\documentclass[preprint,review, 12pt]{elsarticle}
 
\usepackage{amssymb}

\usepackage{lineno}
\usepackage{graphicx}
\usepackage{subfigure}
\usepackage{multirow}
\usepackage{amsmath}
\journal{NIMA}

\begin{document}

\begin{frontmatter}



\title{Beam test results of IHEP-NDL Low Gain Avalanche Detectors(LGAD)}
\author[label1,label2]{S. Xiao} 

\author[label3]{S. Alderweireldt}
\author[labelsinica]{S. Ali}
\author[label3]{C. Allaire}
\author[label5]{C. Agapopoulou}
\author[label6]{N. Atanov}
\author[label1]{M. K. Ayoub}
\author[label4]{G. Barone}
\author[labeluh2c]{D. Benchekroun}
\author[labelsinica]{A. Buzatu}
% ...


\address[label1]{Institute of High Energy Physics, Chinese Academy of Sciences, 19B Yuquan Road, Shijingshan District, Beijing 100049, China}
\address[label2]{University of Chinese Academy of Sciences, 19A Yuquan Road, Shijingshan District, Beijing 100049, China}
\address[label3]{CERN, Esplanade des Particules 1, 1211 Geneva 23 }
\address[labelsinica]{Academia Sinica, Nangang District, Taipei}
\address[label5]{LAL, IN2P3-CNRS and Universit\'{e} Paris Sud, 91898 Orsay Cedex, France}
\address[label6]{Joint Institute for Nuclear Research, Joliot-Curie street 6, Dubna, 141980 Russia}
\address[label4]{Brookhaven National Laboratory (BNL), Upton , NY 11973, U.S.A.}
\address[labeluh2c]{University of Hassan II Casablanca, Casablanca, Morocco}
% ...

\begin{abstract}

To meet the timing resolution requirement of up-coming High Luminosity LHC (HL-LHC), a new detector based on the Low-Gain Avalanche Detector(LGAD), High-Granularity Timing Detector (HGTD), is under intensive research in ATLAS. 
Two types of IHEP-NDL LGADs(BV60 and BV170) for this update is being developed by Institute of High Energy Physics (IHEP) of Chinese Academic of Sciences (CAS) cooperated with Novel Device Laboratory (NDL) of Beijing Normal University and they are now under detailed study. 
These detectors are tested with $5GeV$ electron beam at DESY. A SiPM detector is chosen as a reference detector to get the timing resolution of LGADs. 
The fluctuation of time difference between LGAD and SiPM is extracted by fitting with a Gaussian function. 
Constant fraction discriminator (CFD) method is used to mitigate the effect of time walk. 
The timing resolution of $41 \pm 1 ps$ and $63 \pm 1 ps$ are obtained for BV60 and BV170 respectively. 

\end{abstract}




\end{frontmatter}


\section{Introduction}

Silicon detectors have been widely used in physics experiments to track the positions of particles for decades. Millions of detectors form large scale scientific facilities, such as ATLAS. On these detectors, tracks could be reconstructed for the identification of each collision and corresponding secondary particles. 
The pileup on ATLAS will go up to $\sim200$ after the LHC upgrade. During this upgrade, the luminosity will be increased by a factor of $5$ to $10$. One way to supress these pileup events is to add timing information in the existing 3D tracking to form the 4D tracking. 

\begin{figure}[hbtp]
\centering
  \includegraphics[width=0.5\linewidth]{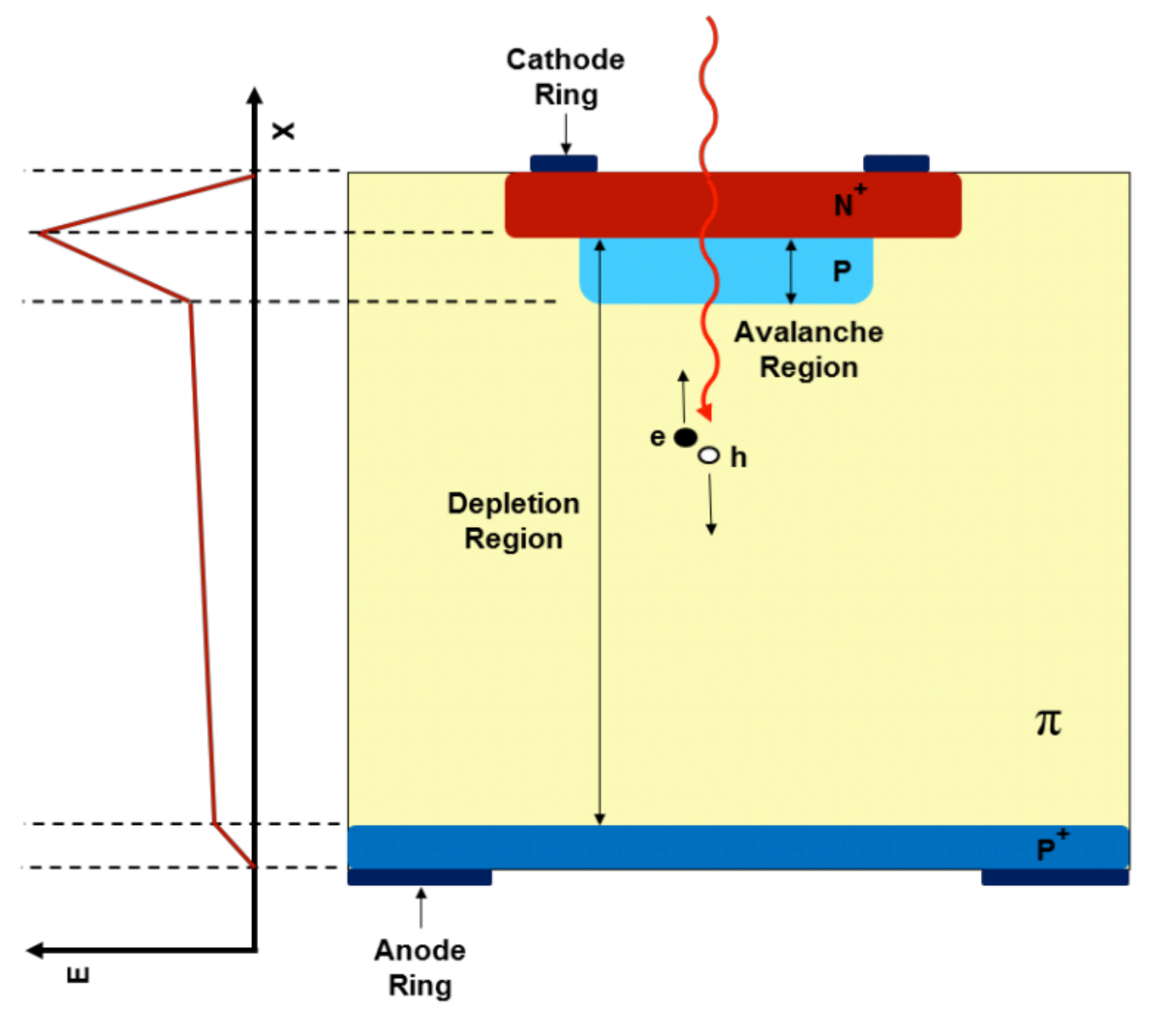}
  \caption{Schematic of an n-on-p LGAD operated under reverse bias voltage.} 
\label{pic_LGAD}
\end{figure}

About  $10 \mu m$ spatial resolution and $\sim$ $45ps$ timing resolution are required for High-Luminosity LHC (HL-LHC).  And the ATLAS high-granularity timing detector (HGTD) is proposed to provide the timing information for the 4D tracking detector. 
A new type of silicon detector is the low-gain avalanche detector (LGAD) \cite{Hartmut, Pellegrini, Gabriele}. A highly doped p layer structure is added to normal PIN diode (Fig.~\ref{pic_LGAD}). 

\begin{figure}[hbtp]
\centering
  \includegraphics[width=0.5\linewidth]{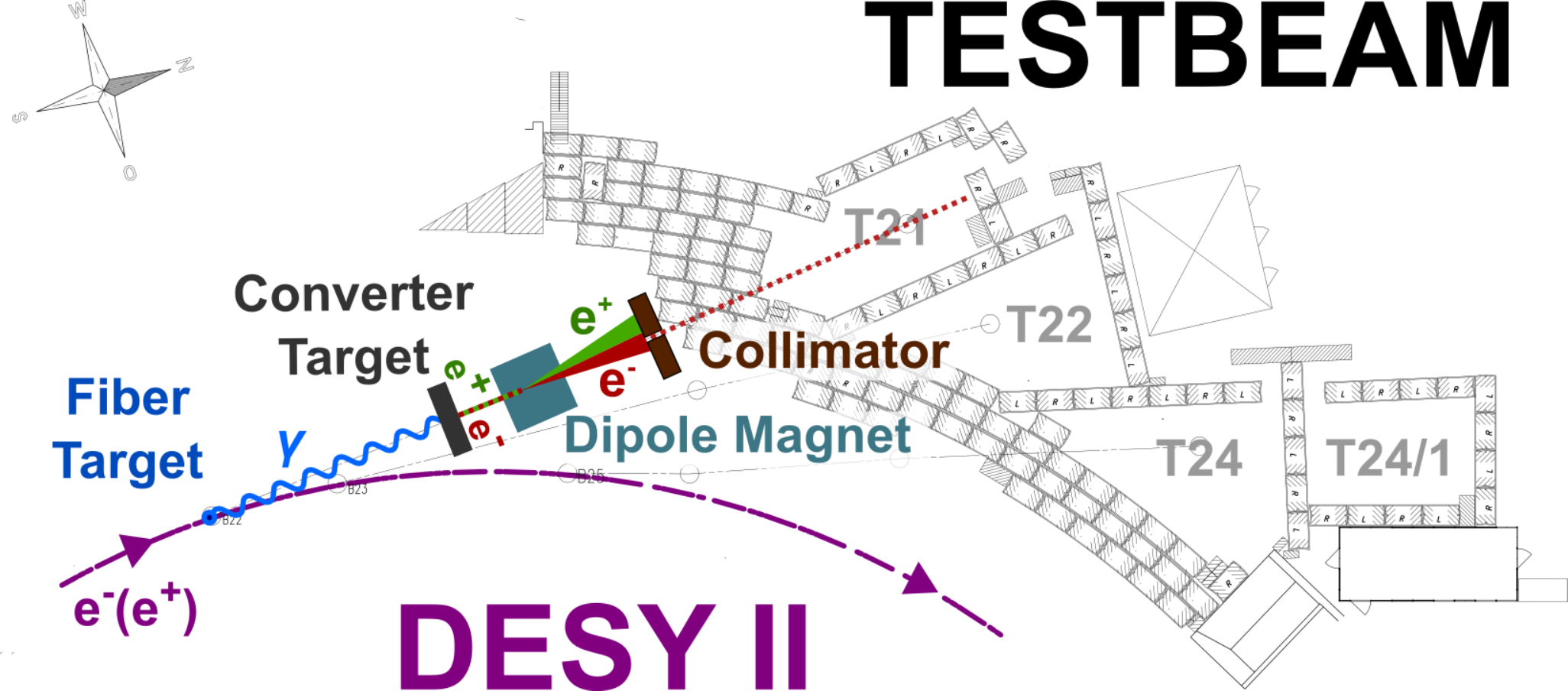}
  \caption{Schematic Layout of a beam test at DESY.} 
\label{pic_layout_tsb}
\end{figure}

To study the properties of IHEP-NDL LGADs, the detectors are tested at DESY using a $5 GeV$ electron beam in room temperature on beam line T22 as
shown in Fig.~\ref{pic_layout_tsb}\cite{Diener}. 
The electron beam is converted bremsstrahlung beams from carbon fibre targets in the electron-positron synchrotron DESY II with up to 1000 particles per $cm^{2}$ and energies from $1$ to $6 GeV$ with an energy spread of $\sim$ 5\% and a divergence of $\sim$ $1mrad$.  

\section{Properties of IHEP-NDL detector}

\begin{figure}[hbtp]
\centering
  \includegraphics[width=0.5\linewidth]{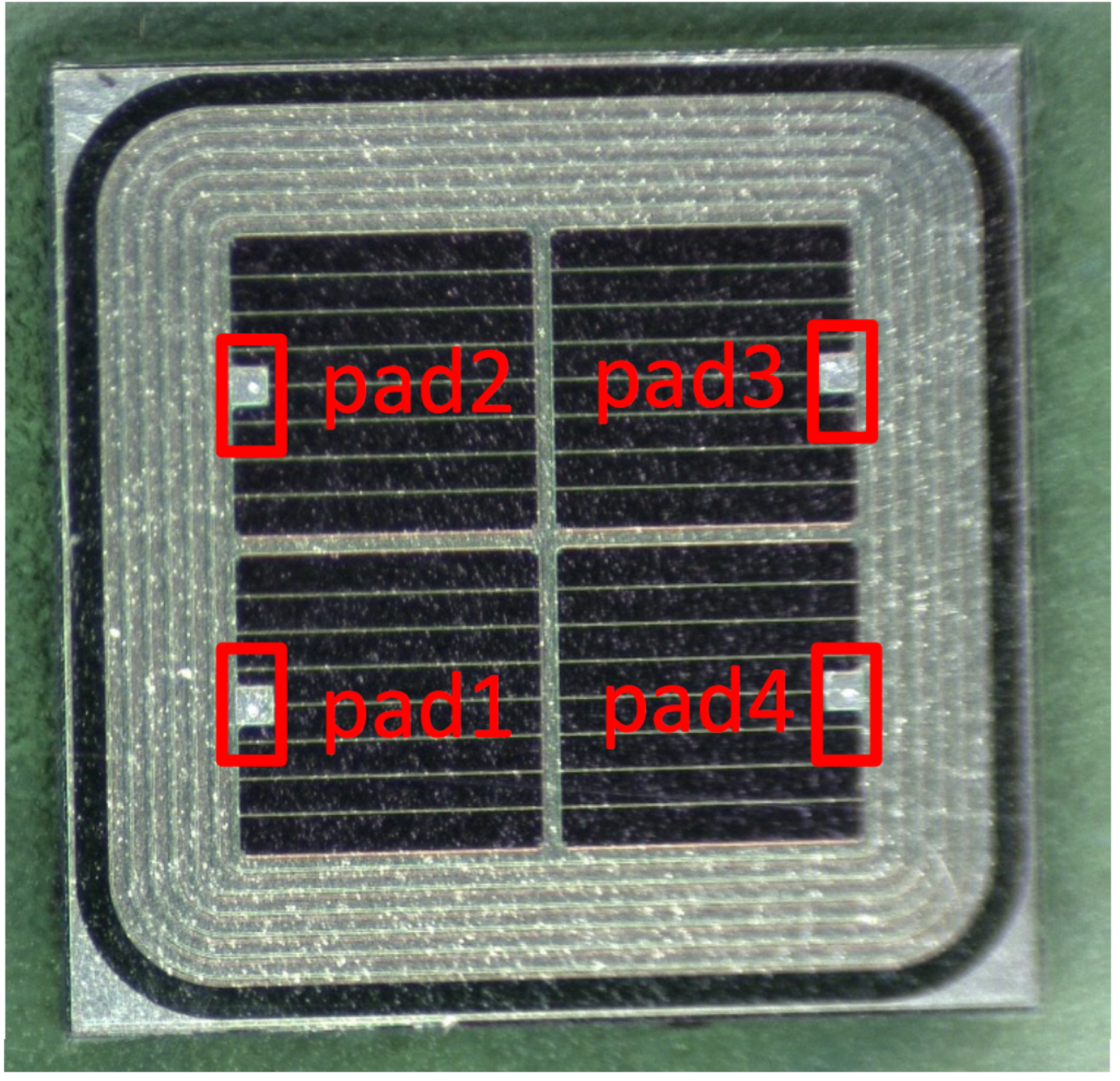}
  \caption{Schematic Layout of a IHEP-NDL detector. $2$ by $2$ pads each detector are fabricated with floating six GRs, no marks for distinguishment. } 
\label{pic_NDL}
\end{figure}

The detectors to be tested, BV60 and BV170, are provided by IHEP-NDL. 
$2$ by $2$ pads each detector are both fabricated with multiple inactive guard rings (GR) (Fig.~\ref{pic_NDL}), four pads behaving almost the same and no marks on the surface could be used to distinguish these pads. 
No electrodes for these six GRs in this first version of IHEP-NDL LGADs are connected. The detector has a size of $ 3.2 \times 3.2mm^{2}$ . 
The physical thickness of detector is $300 \mu m$, while the epitaxial layer thickness is $33 \mu m$.

\begin{figure}[hbtp]
\centering
\subfigure[IV]{
\begin{minipage}[t]{0.45\linewidth}
  \includegraphics[width=1\linewidth]{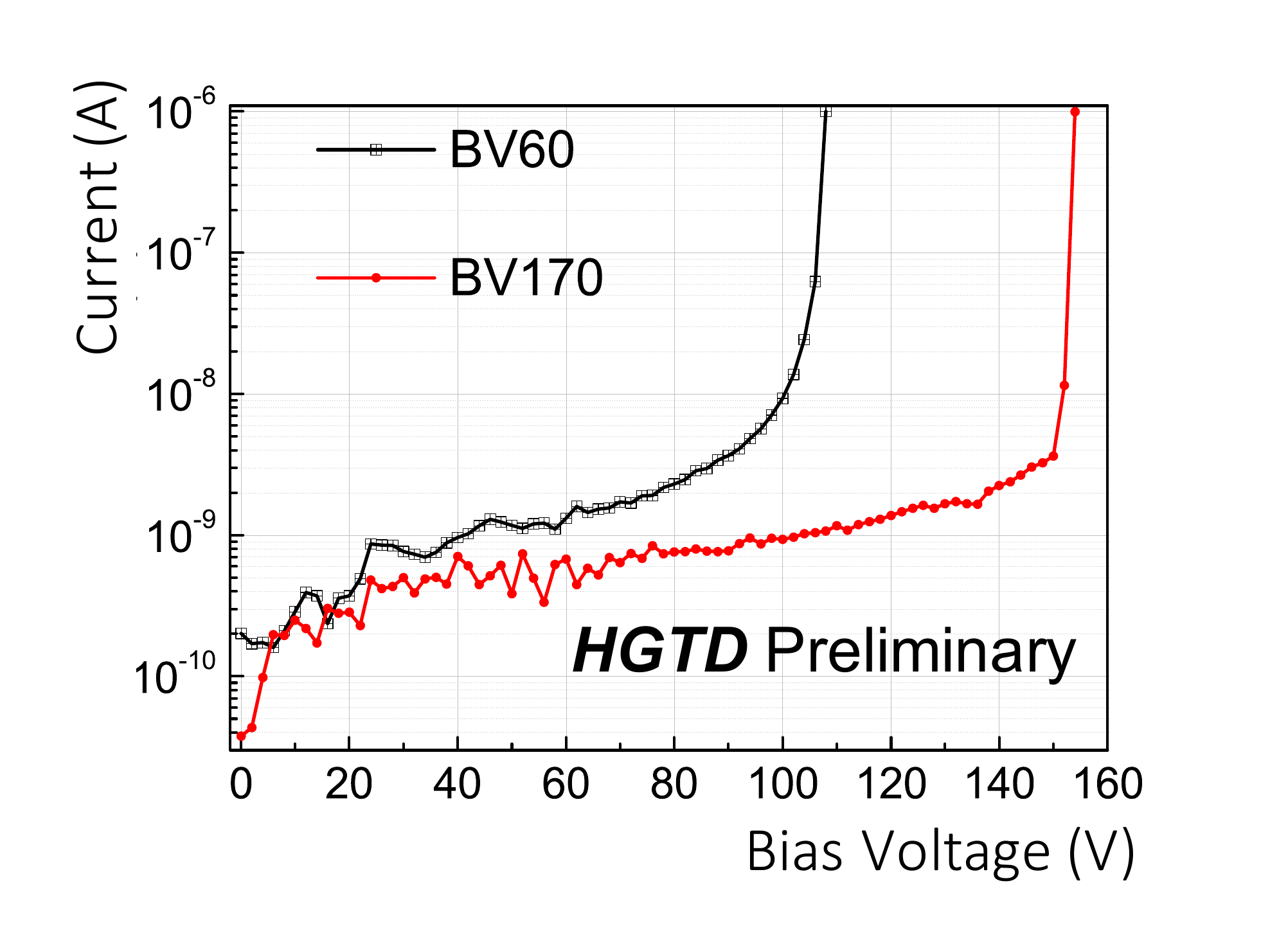}
\end{minipage}
}
\subfigure[$C^{-2}$]{
\begin{minipage}[t]{0.45\linewidth}
  \includegraphics[width=1\linewidth]{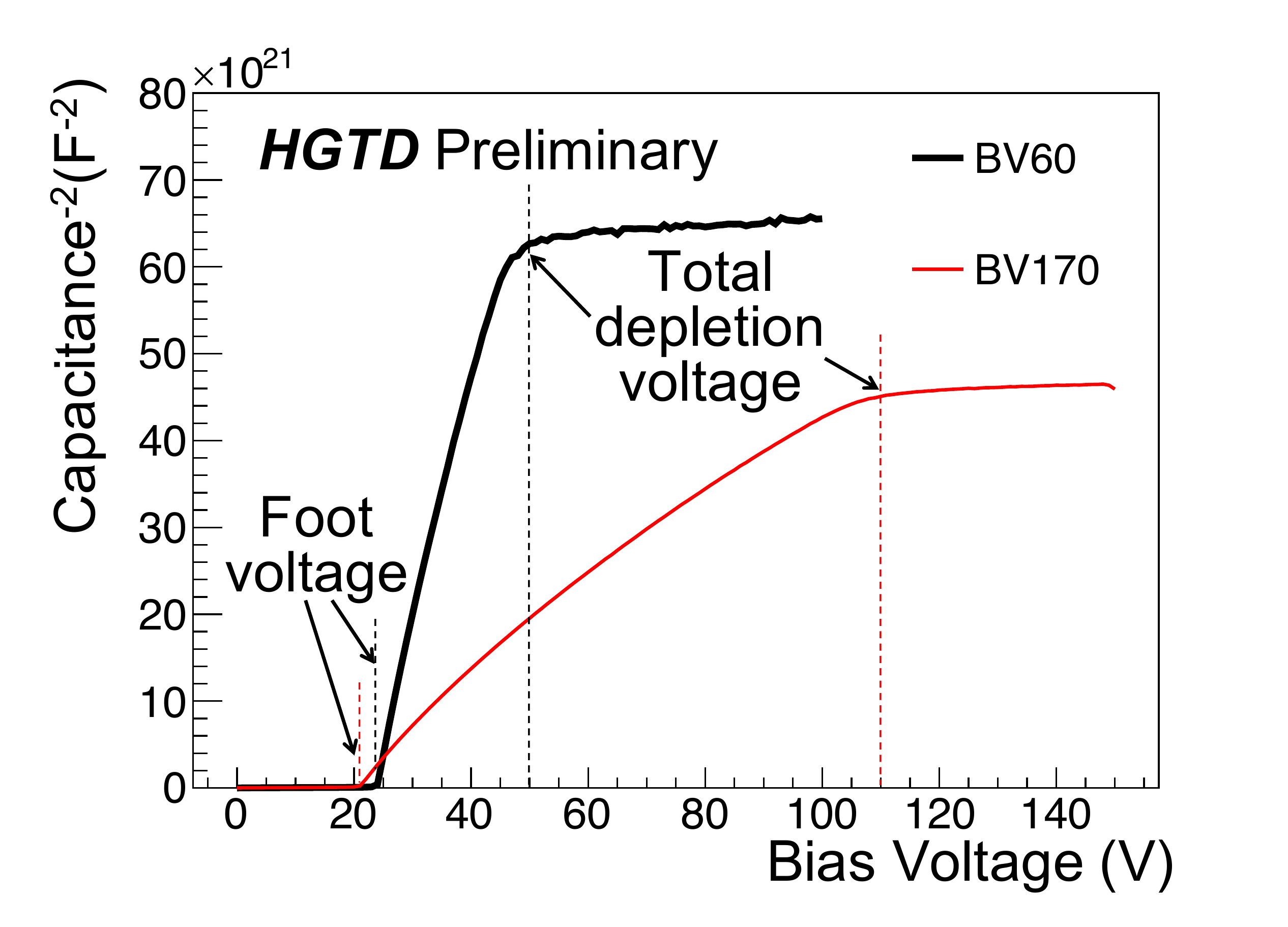}
\end{minipage}
}
\centering
  \caption{The IV curve of IHEP-NDL LGADs and $C^{-2}$ vs bias voltage for IHEP-NDL LGADs.} 
\label{pic_IVCV}
\end{figure}

The Current-Voltage  (I-V) and Capacitance-Voltage  (C-V) measurements are carried out with a probe station at room temperature, GR floating(Fig.~\ref{pic_IVCV}). The reversed bias voltages are applied to the LGADs. Only absolute value of the bias voltage is discussed both in the following text and plots for simplification. Depletion voltage, breakdown voltage and foot voltage could be extracted from the plots. The foot voltage is the bias voltage to completely depleted the gain layer. 
BV60 depletes at $ 50V$ and breaks down at $110V$ , the foot voltage locating at $24V$. BV170 depletes at $110V$ and breaks down at $160V$,  the foot voltage locating at $21V$ . 

The timing resolution of LGADs could be describe as Eq.~\ref{formula_time_reso}. 
\begin{equation}\label{formula_time_reso}
\sigma^2 = \sigma_{Landau~noise}^2+\sigma_{Signal~distortion}^2+\sigma_{Time~walk}^2+\sigma_{TDC}^2+\sigma_{Jitter}^2
\end{equation}

$\sigma_{Landau~noise}$ is effected by the non-uniform energy deposition caused by the beam particles. 
$\sigma_{Signal~distortion}$ represents the uncertainty caused by the non uniform drift velocity and weighting field according to the Ramo-Shockley's theorem \cite{Hartmut}.
$\sigma_{Time~walk}$ is caused by the arrival time shift among pulses with different amplitudes. 
Constant fraction discriminator (CFD) method is used to reduce the effect from time walk. Three cases, $20\%$, $50\%$ and $70\%$, will be discussed. 
$\sigma_{TDC}$ is the contribution of TDC binning, which is usually below 10 ps and therefore it will be neglected. 
$\sigma_{Jitter}$ is from the noise on the signal or electronics and could be estimated by $\sigma_{Jitter}$=N/(dV/dt) $\approx$ t$_{rise}$/(S/N) \cite{Hartmut}. 
The jitter term is evaluated using pico-second laser and fast sampling rate oscilloscope\cite{Yuzhen} to be $10ps$. 

Another important parameter of LGAD is the gain, usually calculating by the collected charges of LGAD and corresponding PIN. All structure should be the same except the gain layer. 
No PINs are fabricated in this batch, so gain measurements will be reported in next papers.

\section{Experimental set-up}
\begin{figure}[hbtp]
\centering
  \includegraphics[width=0.4\linewidth]{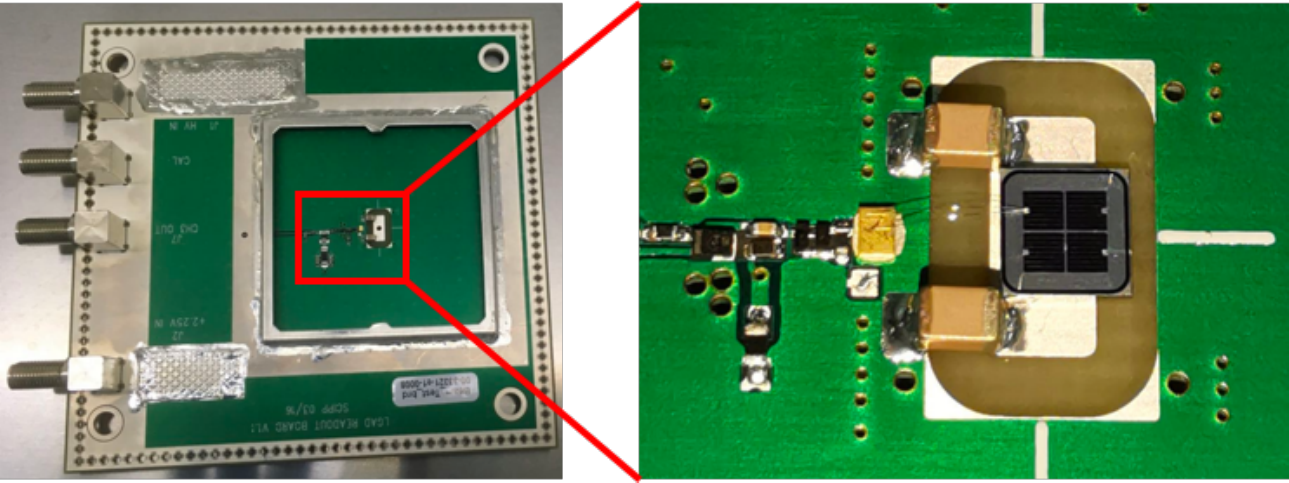}
  \caption{Tested detector is wire bonded to UCSC read-out board in size of $\sim$ $10 cm^{2}$ . A $5G$Hz silicon phosphate is coupled to a $2G$Hz amplifier. } 
\label{pic_wbond}
\end{figure}

\begin{figure}[hbtp]
\centering
  \includegraphics[width=0.5\linewidth]{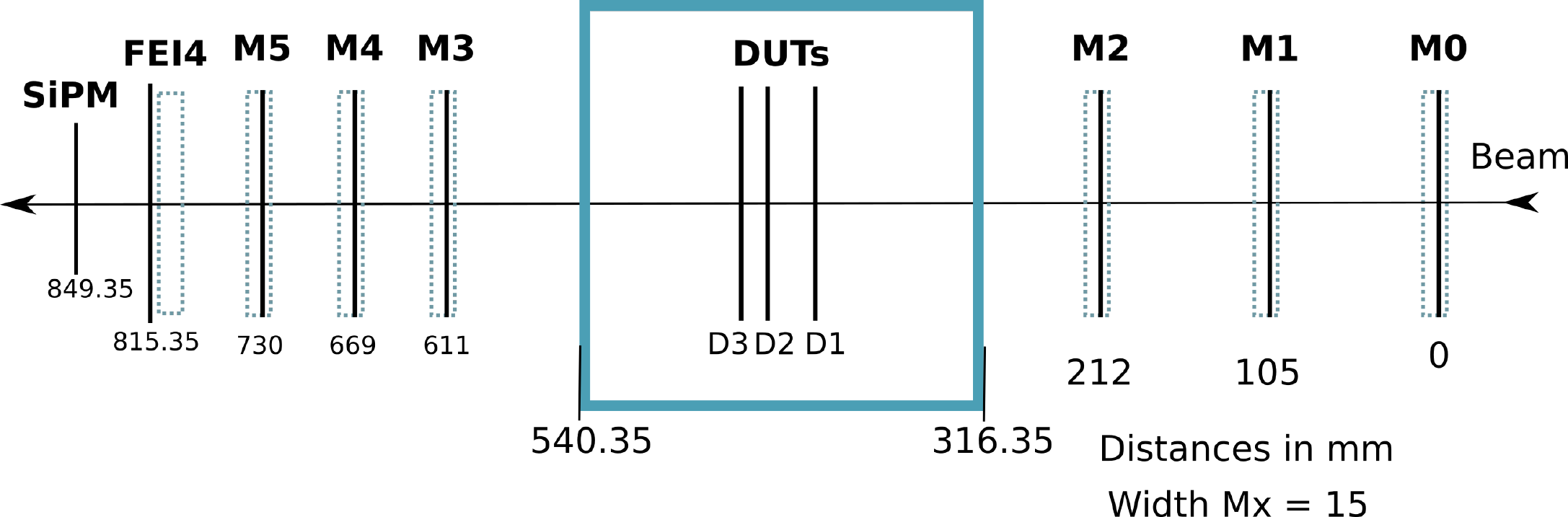}
  \caption{Sketch of positions in the setup.} 
\label{pic_setup}
\end{figure}

As shown in Fig.~\ref{pic_wbond},  the LGAD is wire bonded to a $\sim$ $10 cm$ square read-out board, a $5G$Hz silicon phosphate coupled to a $2G$Hz amplifier \cite{Cartiglia}, developed at the University of California Santa Cruz  (UCSC). The EUDET-type beam telescope \cite{Jansen} is consist of six pixel detector planes equipped with MIMOSA sensors. It is used to acquire native data by Trigger Logic Unit(TLU), cluster event entries and form hits on the telescope planes, align the telescope planes and fit the tracks. MIMOSA planes and FEI4 are fixed as shown in Fig.~\ref{pic_setup}, and DUTs (detectors under test) can be moved according to the location requirements. For BV60 and BV170, DUTs are located $D1 = 403.35 mm$, $D2 = 430.35 mm$ and $D3 = 443.35 mm$ ($\pm 1 mm$). 
Negative voltages are applied to IHEP-NDL LGADs, and a positive voltage to reference SiPM. 
For simplification, only absolute values of bias voltages applied on LGADs will be discussed. $50V$, $90V$ and $130V$ are applied to BV170 while $50V$, $70V$ and $90V$ to BV60. $26.5V$ bias voltage is also applied to the reference SiPM. 

\section{Results}

\subsection{Data analysis}

The oscilloscope records all waveforms generated during the measurement, and the main information of detectors.
The EUDET-type beam telescope \cite{Jansen} is used to acquire the native data.  An analysis framework, PyAna \cite{pyana}, is used to reconstruct the HGTD beam test data. This framework takes the raw oscilloscope data as input and generates a root file containing a tree with various variables as output. The collected charge could be calculated by $Q=I\times t$. For inconstant current, the charges estimated by $Q=\sum_{i}^{n}(I_{i} \times \Delta t)$. $ \Delta t$ is a constant time interval between the data acquisition. 
CFD method is studied by the time during which period the amplitude is higher than a percentage of the maximum amplitude. The time is used only to investigate the timing resolution in the next section.


The amplitude and collected charge is fitted with a Landau function convoluted with a Gaussian function. The most probable value (MPV) is taken as amplitude or collected charge at certain bias voltage. A Gaussian function is used to describe the noise distribution, and the sigma of this fitted Gaussian function is taken as the noise. S/N is calculated using amplitude over noise. 

\begin{figure}[hbtp]
\centering
\subfigure[Amplitude]{
\begin{minipage}[t]{0.47\linewidth}
  \includegraphics[width=\linewidth]{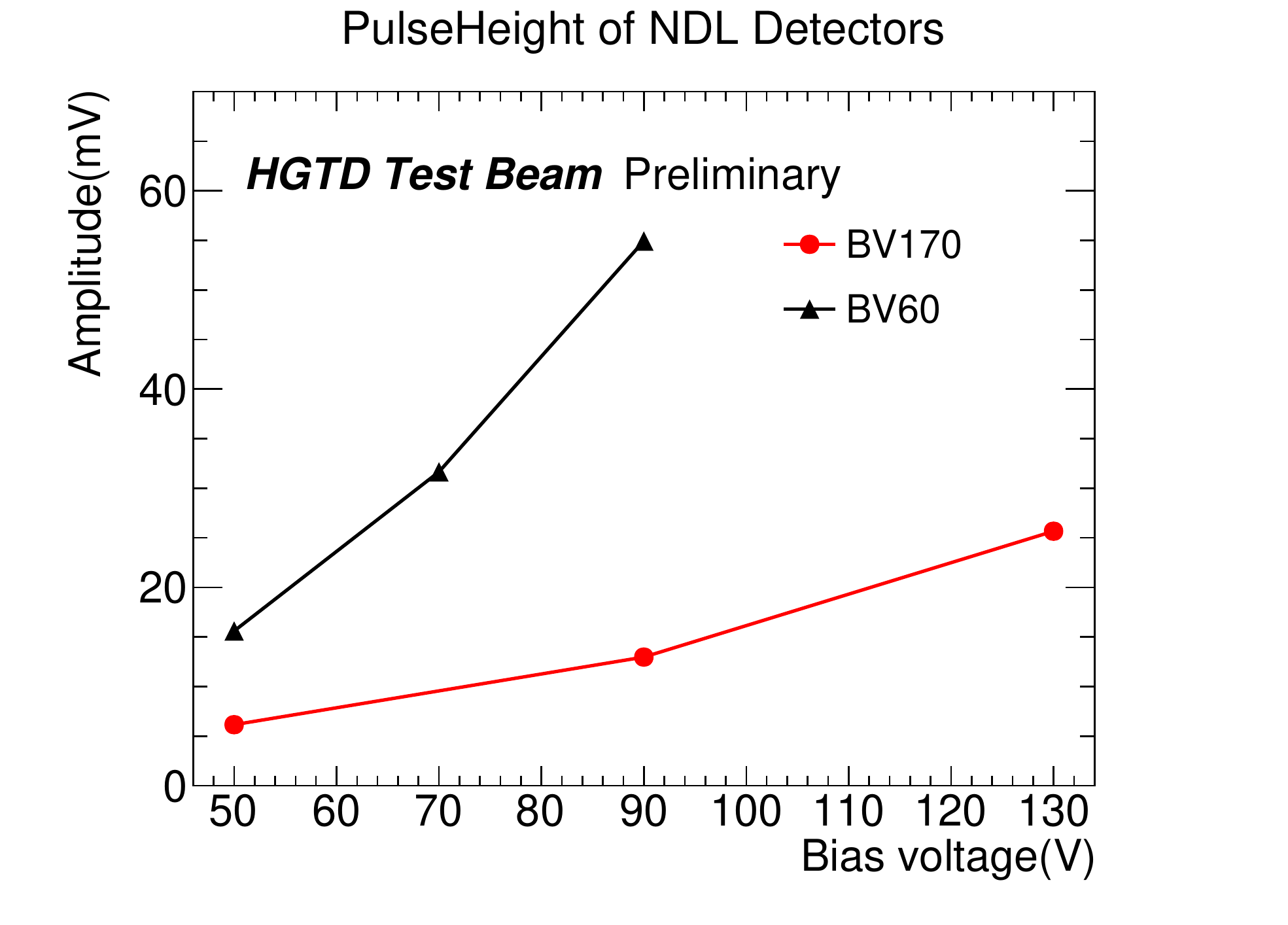}
\end{minipage}
}
\subfigure[Noise]{
\begin{minipage}[t]{0.47\linewidth}
  \includegraphics[width=\linewidth]{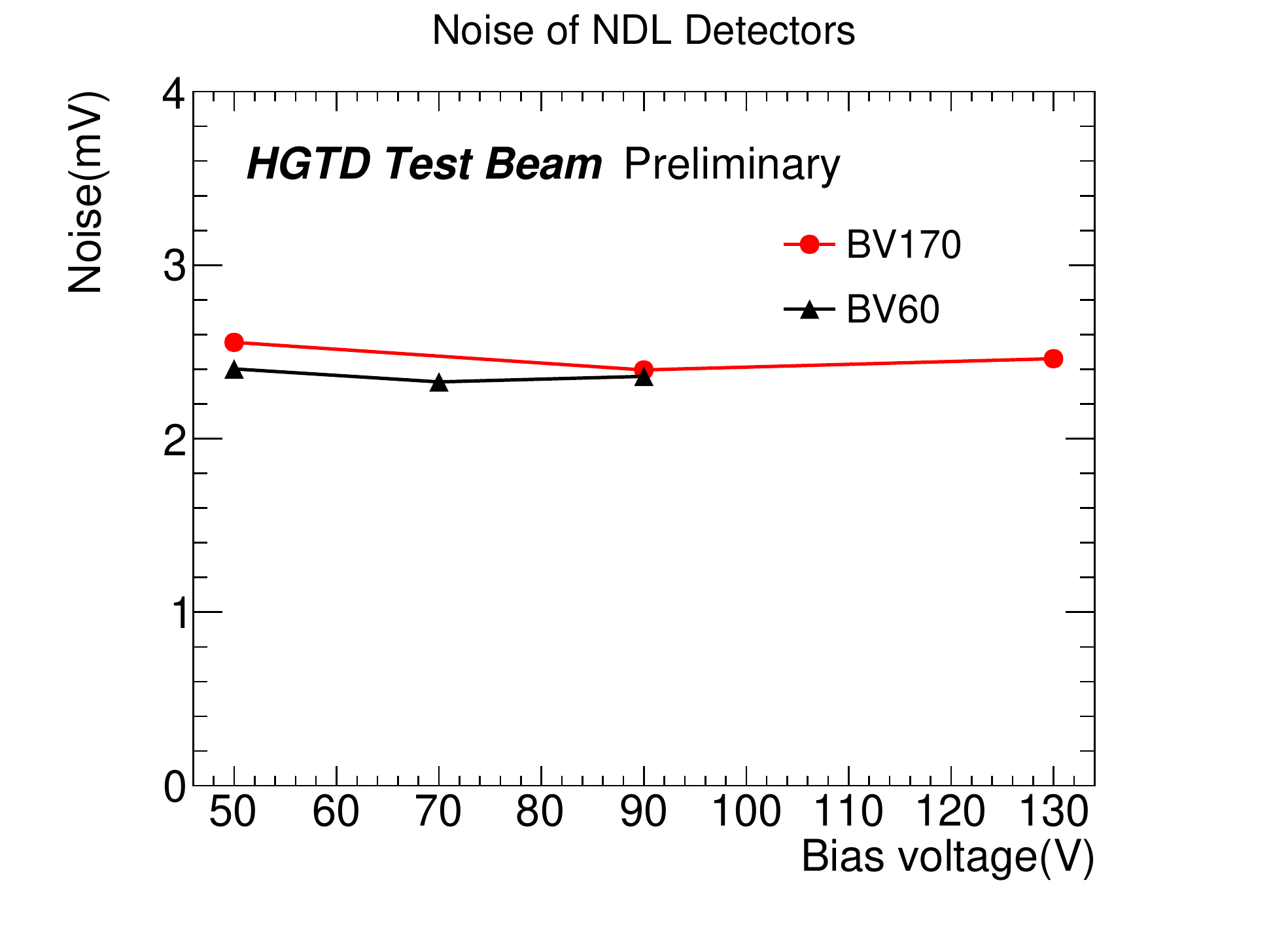}
\end{minipage}
}
\subfigure[S/N]{
\begin{minipage}[t]{0.47\linewidth}
  \includegraphics[width=\linewidth]{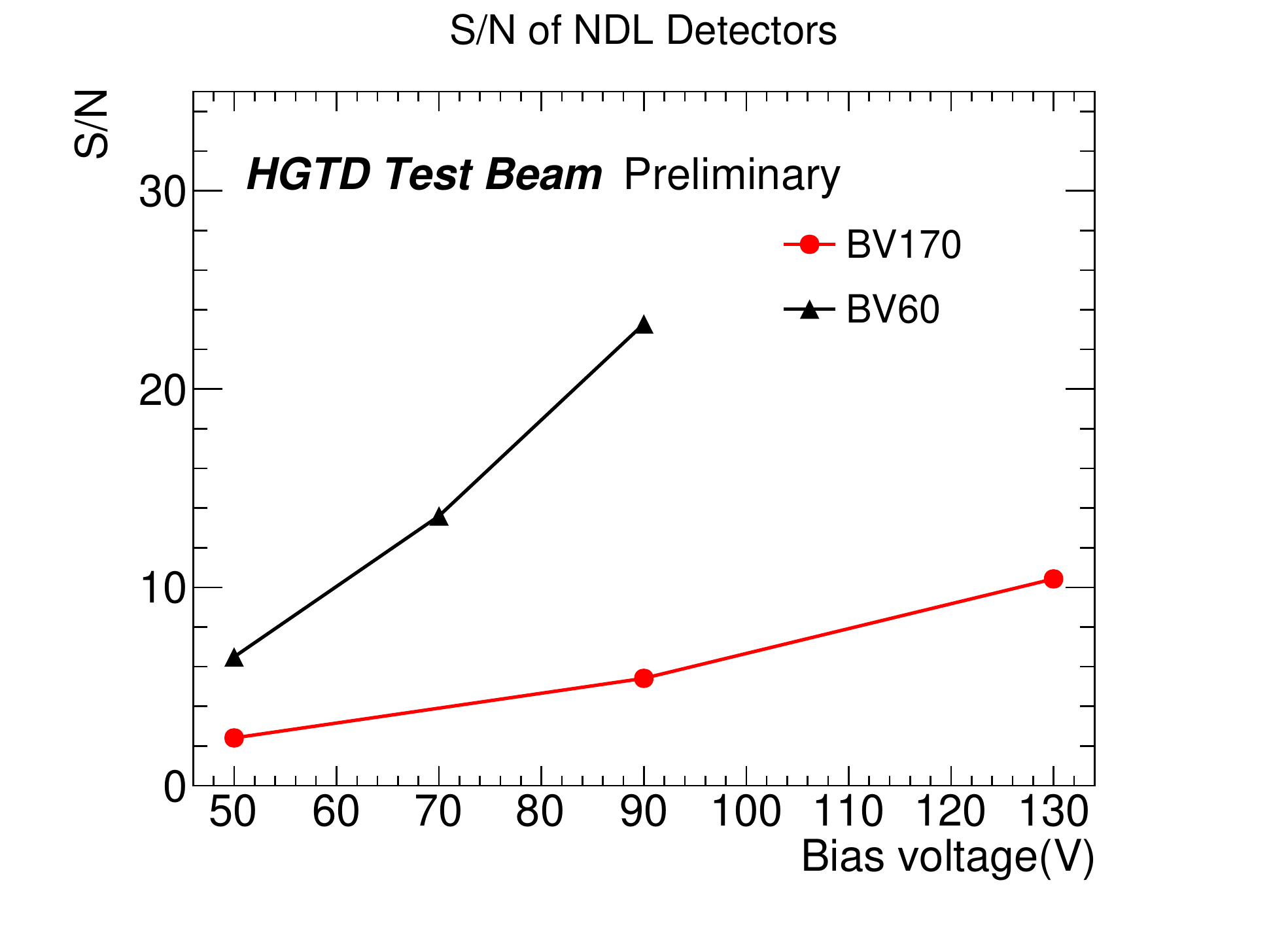}
\end{minipage}
}
\subfigure[Charge]{
\begin{minipage}[t]{0.47\linewidth}
  \includegraphics[width=\linewidth]{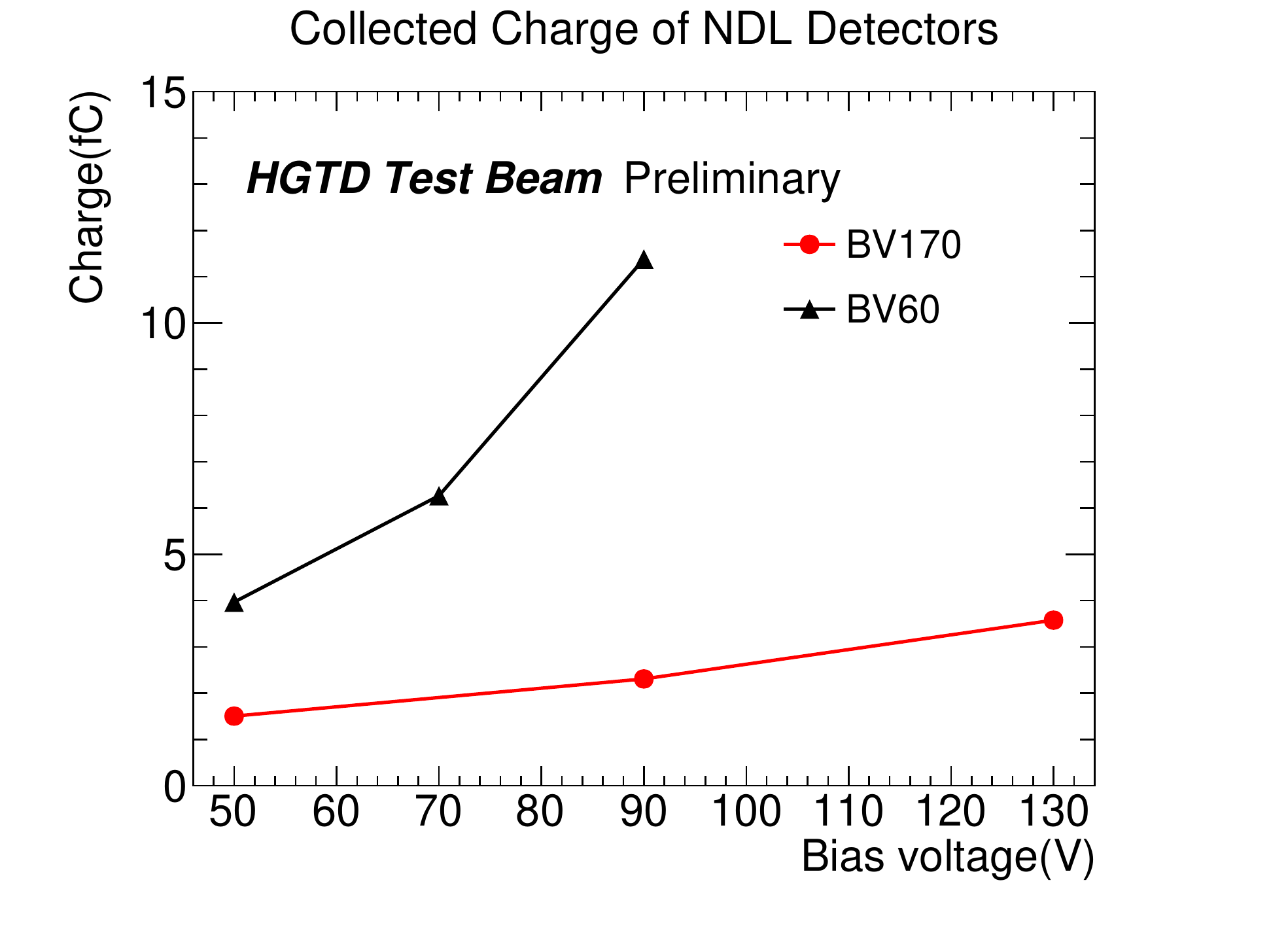}
\end{minipage}
}
\centering
\caption{ (a) Amplitude distribution for BV60 and BV170. 
 (b) Noise distribution for BV60 and BV170.  (c) S/N distribution for BV60 and BV170, calculating using amplitude over noise. (d) Collected charge for BV60 and BV170. BV60 collects more charges than $4 fC$. }
\label{pic_results}
\end{figure}

The performance of these two detectors are shown in Fig~\ref{pic_results}. The higher bias voltage leads to higher amplitude and more collected charges. 
The noise remains almost the same at different bias voltages, resulting in S/N  a similar trend with amplitude. 
BV60 collects more charges than $4 fC$, but BV170 fails to do so at tested voltages. But the properties of BV170 at higher bias voltages are still worth being studied, for example, above $150V$. $130V$ is somehow too far from the breakdown voltage, which is probably the main reason causing the bad behavior of BV170. 

\subsection{Timing resolution}
As mentioned above, SiPM is taken as a reference detector in order to calculate the timing resolutions. 
Three different combinations can be obtained with BV60, BV170 and SiPM.
And these time differences(based on different CFD valus) can be described well with Gaussian functions. The parameter sigma of the fitted Gaussian function could be extracted as combined resolutions for the three systems. 


By assuming that the two detectors in a combination are non-correlated and taking the system containing BV60 and SiPM as an example, the timing resolution of BV60 could be calculated as

\begin{equation}
\sigma_{1}=\sqrt{\sigma_{13}^{2}-\sigma_{3}^{2}}
\end{equation}
taking $\sigma_{1}$ and $\delta_{1}$ as timing resolution and uncertainty of BV60,  
$\sigma_{3}$ and $\delta_{3}$ as SiPM, with an uncertainty
\begin{equation}
\delta_{1}=\sqrt{\frac{ (\sigma_{13}\delta_{13})^{2}+ (\sigma_{3}\delta_{3})^{2}}{\sigma_{13}^{2}-\sigma_{3}^{2}}}
\end{equation}

Same to other two cases, BV60 and BV170 ($\sigma_{2}$ with $\delta_{2}$), BV170 and SiPM. Then calculating for the $\sigma$ 
\begin{equation}
\begin{split}
\sigma_{1}=\sqrt{\frac{\sigma_{21}^{2}+\sigma_{13}^{2}-\sigma_{32}^{2}}{2}}, \\
\sigma_{2}=\sqrt{\frac{\sigma_{21}^{2}-\sigma_{13}^{2}+\sigma_{32}^{2}}{2}},\\
\sigma_{3}=\sqrt{\frac{-\sigma_{21}^{2}+\sigma_{13}^{2}+\sigma_{32}^{2}}{2}}
\end{split}
\end{equation}

with uncertainty 
\begin{equation}
\begin{split}
\delta_{1}=\frac{\sqrt{ (\sigma_{21}\delta_{21})^{2}+ (\sigma_{13}\delta_{13})^{2}+ (\sigma_{32}\delta_{32})^{2}}}{2\sigma_{1}}, \\
\delta_{2}=\frac{\sqrt{ (\sigma_{21}\delta_{21})^{2}+ (\sigma_{13}\delta_{13})^{2}+ (\sigma_{32}\delta_{32})^{2}}}{2\sigma_{2}}, \\
\delta_{3}=\frac{\sqrt{ (\sigma_{21}\delta_{21})^{2}+ (\sigma_{13}\delta_{13})^{2}+ (\sigma_{32}\delta_{32})^{2}}}{2\sigma_{3}}
\end{split}
\end{equation}

\begin{figure}[hbtp]
\centering
  \includegraphics[width=0.5\linewidth]{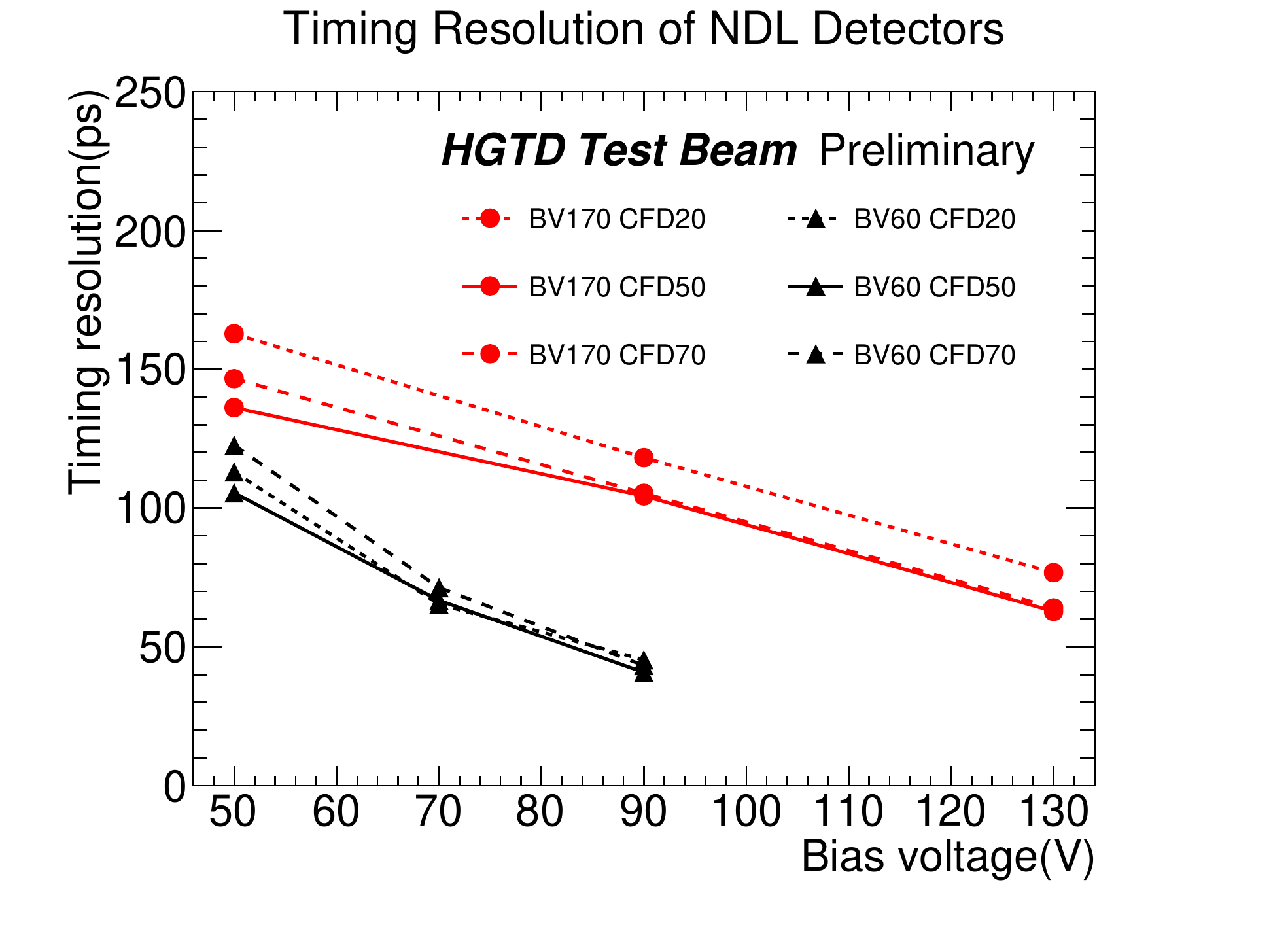}
  \caption{Timing resolution for BV60 and BV170 as function of voltage. The timing resolutions of these two detectors are getting better as bias voltage goes up. The CFD50 leads to smaller timing resolutions than CFD20 and CFD70.} 
\label{pic_time_res}
\end{figure}

Timing resolutions for BV60 and BV170 are shown in Fig~\ref{pic_time_res}. The timing resolutions of these two detectors are getting better as bias voltage goes up. The effect from CFD values is shown clearly. The CFD50, black solid line(color online) for BV60 and red solid line for BV170, leads to smaller timing resolutions than CFD20 and CFD70. 
All the timing resolutions of BV60, BV170 and SiPM are listed in Table~\ref{timing res 101}, Table~\ref{timing res 102} and Table~\ref{timing res 103} for three different bias voltages. 

\begin{table}[hbtp]
\caption{Timing resolution for detectors at bias voltage: $50V$ for BV60, $50V$ for BV170, and $26.5V$ for SiPM.} 
\label{timing res 101}
\small
\begin{center}
\renewcommand{\arraystretch}{1.2}
\begin{tabular}{lcccc}
\hline
\hline
CFD[\%]		&BV60[ps] 	 				& BV170[ps]  	 				& SiPM[ps] 	 		\\ 
\hline 
	20		&  	 113 $\pm$ 6		& 	163 $\pm$ 4		&	51 $\pm$ 13			\\ 
	50		&  	 105 $\pm$ 5		& 	136 $\pm$ 4		&	60 $\pm$ 8			\\ 
	70		&  	 122 $\pm$ 5		& 	147 $\pm$ 4		& 51 $\pm$ 11			\\ 
\hline
\hline
\end{tabular}
\end{center}
\end{table}

\begin{table}[hbtp]
\caption{Timing resolution for detectors at bias voltage: $70V$ for BV60, $90V$ for BV170, and $26.5V$ for SiPM.} 
\label{timing res 102}
\small
\begin{center}
\renewcommand{\arraystretch}{1.2}
\begin{tabular}{lcccc}
\hline
\hline
CFD[\%]		&BV60[ps]  	 				& BV170[ps]  	 				& SiPM[ps] 	 		\\ 
\hline 
	20		&  	 65 $\pm$ 2		& 	118 $\pm$ 1		&	74 $\pm$ 2			\\ 
	50		&  	 67 $\pm$ 2		& 	104 $\pm$ 1		&	65 $\pm$ 2			\\ 
	70		&  	 71 $\pm$ 2		& 	105 $\pm$ 1		& 65 $\pm$ 2			\\ 
\hline
\hline
\end{tabular}
\end{center}
\end{table}

\begin{table}[hbtp]
\caption{Timing resolution for detectors at bias voltage: $90V$ for BV60, $130V$ for BV170, and $26.5V$ for SiPM.} 
\label{timing res 103}
\small
\begin{center}
\renewcommand{\arraystretch}{1.2}
\begin{tabular}{lcccc}
\hline
\hline
CFD[\%]		&BV60[ps]  	 				& BV170[ps]  	 				& SiPM[ps] 	 		\\ 
\hline 
	20		&  	 45 $\pm$ 1		& 	77 $\pm$ 1		&	73 $\pm$ 1			\\ 
	50		&  	 41 $\pm$ 1		& 	63 $\pm$ 1		&	70 $\pm$ 1			\\ 
	70		&  	 43 $\pm$ 1		& 	65 $\pm$ 1		& 70 $\pm$ 1			\\ 
\hline
\hline
\end{tabular}
\end{center}
\end{table}


\section{Conclusion}

IV and CV tests have been carried out on BV60 and BV170. The voltage points applied on the beam test are determined according to the depletion voltages and breakdown voltages. The amplitude of BV60 is higher than that of BV170, while they have similar noise, which resulting in higher S/N than BV170. The BV60 could collect more charges than $4fC$. More investigation of BV170 on collected charge at higher bias voltages near to breakdown voltage are still worth being studied, for example, $150V$ or $155V$. 
Different CFD values do effect on timing resolution, CFD50 leading to the best results comparing with CFD20 and CFD70. The timing resolution of BV60 is $41\pm 1ps$, and BV170 is $63\pm 1ps $. 

The reference SiPM has been used in several beam tests at CERN and DESY. Long-term exposure to particles results in a degraded performance, $70\pm 1ps$ at $26.5V$. And this could be improved by using a new SiPM. 



 Irradiation hardness is another issue that matters as well. 
IHEP-NDL LGADs have been sent to China Institute of Atomic Energy (CIAE) and Cyclotron and Radioisotope Center (CYRIC) for proton irradiation up to $2.5\times 10^{15} neq/cm^{2}$. IV, CV, gain measurement and beam test will all be carried out on these irradiated LGADs. Detailed researches can be expected in near future.

\section*{Acknowledgement}

The measurements leading to these results have been performed at the beam test Facility at DESY Hamburg (Germany), a member of the Helmholtz Association (HGF). 
Thanks to Beijing Normal University for the detector production, and DESY for the beam test setup and shifts. 
Thanks all colleagues involved in these processes. This work was supported by the National Natural Science Foundation of China (No. 11961141014), the State Key Laboratory of Particle  Detection and Electronics (SKLPDE-ZZ-202001), and the Hundred Talent Program of the Chinese Academy of Sciences (Y6291150K2).

\section*{References}

\bibliographystyle{unsrt}

\end{document}